\documentclass[10pt,conference]{IEEEtran}
\pdfoutput=1
\IEEEoverridecommandlockouts
\usepackage{cite}
\usepackage{amsmath,amssymb,amsfonts}
\usepackage{algorithmic}
\usepackage{graphicx}
\usepackage{textcomp}
\usepackage{xcolor}
\usepackage{framed}
\usepackage{svg}
\usepackage{epsfig}
\usepackage{multirow}
\usepackage{graphicx}  
\usepackage{epstopdf}
  
\usepackage{footnote}
\makesavenoteenv{tabular}

\def\BibTeX{{\rm B\kern-.05em{\sc i\kern-.025em b}\kern-.08em
    T\kern-.1667em\lower.7ex\hbox{E}\kern-.125emX}}
\begin{document}

\title{Predicting Community Smells' Occurrence on Individual Developers by Sentiments\\
\thanks{Corresponding Authors: Zhiqing Shao and Guisheng Fan.}
}

\author{

\IEEEauthorblockN{Zijie Huang\textsuperscript{†}, Zhiqing Shao\textsuperscript{†}, Guisheng Fan\textsuperscript{†}, Jianhua Gao\textsuperscript{§}, Ziyi Zhou\textsuperscript{†}, Kang Yang\textsuperscript{†}, Xingguang Yang\textsuperscript{†}}
\IEEEauthorblockA{
\textit{\textsuperscript{†}Department of Computer Science and Engineering, East China University of Science and Technology, Shanghai, China}\\
\textit{\textsuperscript{§}Department of Computer Science and Technology, Shanghai Normal University, Shanghai, China}\\
hzj@mail.ecust.edu.cn, \{zshao, gsfan\}@ecust.edu.cn,}\ jhgao@shnu.edu.cn, zhouziyi@mail.ecust.edu.cn,  \{15921709583, xingguang2955\}@163.com
}


\maketitle

\begin{abstract}
Community smells appear in sub-optimal software development community structures, causing unforeseen additional project costs, e.g., lower productivity and more technical debt. Previous studies analyzed and predicted community smells in the granularity of community sub-groups using socio-technical factors. However, refactoring such smells requires the effort of developers individually. To eliminate them, supportive measures for every developer should be constructed according to their motifs and working states. Recent work revealed developers' personalities could influence community smells' variation, and their sentiments could impact productivity. Thus, sentiments could be evaluated to predict community smells' occurrence on them. To this aim, this paper builds a developer-oriented and sentiment-aware community smell prediction model considering 3 smells such as Organizational Silo, Lone Wolf, and Bottleneck. Furthermore, it also predicts if a developer quitted the community after being affected by any smell. The proposed model achieves cross- and within-project prediction F-Measure ranging from 76\% to 93\%. Research also reveals 6 sentimental features having stronger predictive power compared with activeness metrics. Imperative and indicative expressions, politeness, and several emotions are the most powerful predictors. Finally, we test statistically the mean and distribution of sentimental features. Based on our findings, we suggest developers should communicate in a straightforward and polite way.

\end{abstract}


\section{Introduction}
Software engineering is a complex social activity that incorporates both communication and collaboration of developers and other stakeholders\cite{PalombaCSDetect}. The social and socio-technical features of development communities, along with the purely technical ones\cite{PalombaCSDetect}, have already been applied to predict and analyze software reliability\cite{defect_prediction} and maintainability\cite{PalombaCSImpact}. Inspired by the conception of code smell\cite{fowler1999refactoring}, Palomba \textit{et al.} \cite{PalombaCSDetect} coined the term “community smell” to describe unhealthy organizational structure of the software development communities causing social debt\cite{Social_Debt,SocialDebt2}. Social debt, much like technical debt\cite{TechdEBT} in terms of the impact on software maintenance, refers to unforeseen project costs connected to the presence of non-cohesive development communities having communication or collaboration issues. Such smells may not be the direct cause of software defects, however, they were proved to be refactoring preventers\cite{PalombaCSImpact}. 

Community smells were evaluated and predicted at the granularity of community sub-groups \cite{PalombaCSDetect, PalombaCSImpact, PalombaCSPredict, PalombaMasters, KBS, ICGSE}. Differently, restructuring smelly communities rely on social efforts of every community member\cite{PalombaCSRefactoring}. Empirical guidelines of refactoring community smells, including monitoring, mentoring, planning, exercising, were proposed to shepherd the community \cite{PalombaCSRefactoring,PalombaCSShepherding,PalombaCSStats}. 

However, open source software development is known for its distributed and egalitarian nature\cite{oss}. Meanwhile, developers focus on primarily their working state and current tasks rather than other  issues\cite{CONTEXT_BASED_REFACTORING_ICPC_2017, PalombaMSR, CONTEXT_DEV1,CONTEXT_DEV_2}. Different from the situations in hierarchical and centralized organizations, the orders and arrangements of community shepherds (\textit{e.g.}, core members\cite{linus}, architects\cite{PalombaCSShepherding}) may not be strictly executed and followed\cite{oss}. Moreover, shepherds may be leaving\cite{linus}, or even missing\cite{unmaintained} from communities. Consequently, general guidelines of software quality that lacked developer-oriented adaptive approaches may be ineffective in practice\cite{PalombaMSR}. 

Since community smells root in developers' motifs \cite{PalombaCSImpact}, research showed personalities could also influence community smells\cite{PalombaCSStats}. Sentiment is an expression of personality\cite{personality}, which can significantly affect the quality of work\cite{Ortu14MSREmotions,Ortu15MSRBullies,Ortu15Politeness,Ortu16MSRVAD,eeg}, and they were proved to have impacts on a variety of tasks including issue reopening\cite{Cheruvelil19}, commit changes\cite{sanerbug,Huq}, code refactoring\cite{Singh_APSEC17_TOOL_NEGATIVE_SENTIMENT}, and security\cite{security}. Previous research suggested sentiment-aware classifiers should be built to support developers' work\cite{eeg}.

To the best of our knowledge, there lacks a community smell prediction model designed individually for developers. Moreover, existing quantitative research focused on analyzing the impact of community smells on code quality rather than social representations of developers. Thus, such models cannot support developers to prevent smell from occurring. As a result, we intend to improve prediction approaches and refactoring guidelines of community smells in the granularity of developers individually by involving their sentiments. 

This paper builds a developer-oriented and sentiment-aware community smell prediction model. We develop machine learners upon a developer sentiment dataset\cite{Ortu16MSRDataset} to predict if a developer is affected by 3 common community smells, such as \textit{Lone Wolf}, \textit{Organization Silo} and \textit{Bottleneck}\cite{PalombaCSDetect}. Furthermore, the model also predicts if a developer is a \textit{Smelly Quitter}\cite{defect_prediction}, \textit{i.e.}, quitted the community after being affected by any community smell concerned. Finally, discussions are made on the difference between smelly and non-smelly developers. We analyze the significance of the difference in sentimental features' distributions as well as its effect size. 
 
The main contributions of the papers are listed as follows:

(1) To our knowledge, the first machine-learning based approach to integrate developer sentiments into community smell prediction, as well as the first work to predict community smells' occurrence on developers individually. The proposed model achieved a cross-project prediction performance of F-Measure ranging from 84\% to 93\%, and a within-project performance of F-Measure ranging from 76\% to 91\%;

(2) We investigate features having the most predictive power. We discover that 6 sentimental features, \textit{i.e.}, imperative and indicative expressions, politeness, and several emotions, are stronger predictors than the activeness metrics.

(3) We evaluate statistically the relationship between community smell and sentiments, which lead us to the conclusion that developers should communicate in a straightforward and polite way to ensure community healthiness;

(4) We provide an online appendix \cite{replication} with the generated dataset to replicate and extend our work.

The rest of this paper is organized as follows: In Section II we summarize related
literature. Section III presents how we construct our dataset, while Section IV outlines the settings and research questions, as well as the concerned evaluation metrics. In Section V we discuss the results of
this study, while Section VI overviews the threats to the validity of
the study and our effort to cope with them. Finally, Section VII concludes
the paper and describes future research.

\section{Related Work}
This section describes researches related to two aspects of this paper, \textit{i.e.}, community smell and developer sentiment.
\subsection{Community Smell}

Tamburri, Palomba \textit{et al.} contributed a series of researches \cite{PalombaCSDetect, PalombaCSImpact, PalombaCSPredict, PalombaMasters, PalombaCSDiversity, PalombaCommunityPattern, PalombaCSRefactoring, PalombaCSStats} concerning the definition\cite{PalombaCSDetect, PalombaMasters}, detection\cite{PalombaCSDetect}, diffuseness\cite{PalombaCSDetect, PalombaCSDiversity}, and variability\cite{PalombaCSStats} of community smells, as well as their impact on software maintainability\cite{PalombaCSImpact}. The authors treat community smells as patterns of motifs over collaboration and communication graphs, and they implemented a detection tool called \textsc{Codeface4Smells} by extending a socio-technical analysis tool called \textsc{Codeface}. The tool accepted the input of development mailing list and software repository history information to detect 5 community smells, namely \textit{Organisational Silo}, \textit{Black Cloud}, \textit{Lone Wolf}, \textit{Bottleneck} and \textit{Prima Donnas}. The authors also validated qualitatively\cite{PalombaCSDetect} the acceptance of the detection result, and they discovered the results were all true positives\cite{PalombaCSPredict}. Furthermore, the authors observed that community smells could be the preventers of refactoring, and they also intensify code smells continuously\cite{PalombaCSImpact}. Alternatively, the authors also discovered the associations between community patterns\cite{PalombaCommunityPattern}, \textit{e.g.}, \textit{Formal Group}, and  \textit{Informal Network}, to specific community smells. 

In terms of analyzing community smell in the granularity of developers individually, Palomba \textit{et al.} also provided practitioners with refactoring suggestions and frameworks\cite{PalombaCSRefactoring}. In a recent study\cite{PalombaCSStats}, the authors pointed out that communicability is important to prevent community smells, and developers' personality plays an important role in producing smells. Alternatively, Ahammed \textit{et al.} \cite{ahmed} made an exploratory study on 7 Apache projects and discovered the commit activities of developers correlate with their involvement of \textit{Missing Links} (\textit{Lone Wolf}) community smell.

The prediction of community smells has also been actively studied by the research community. Palomba and Tamburri \cite{PalombaCSPredict} built a state-of-the-art model to predict community smells' emergence on within- and cross-project scenarios. The research also pointed out socio-technical congruence, communicability and turnover-related metrics are the most powerful predictors. Almarimi \textit{et al.}\cite{KBS,ICGSE} combined Ensemble Classifier Chain (ECC) and Genetic Programming (GP) techniques to predict the emergence of community smells, and they also introduced 4 novel community smells, \textit{i.e.}, \textit{Solution Defiance}, \textit{Sharing Villainy}, \textit{Organizational Skirmish}, and \textit{Truck Factor Smell}. 

The major differences of this work to the above-mentioned smell prediction papers are: (1) we predict the occurrence of smells in the granularity of developers individually rather than sub-groups or communities; (2) we involve developers' sentiment features to build predictors, and we assess their predictive power and statistical characteristics in software projects, which were not covered in previous researches. 

\subsection{Developer Sentiment}


Ortu \textit{et al.} \cite{Ortu14MSREmotions,Ortu15MSRBullies,Ortu15Politeness,Ortu16MSRVAD,Ortu16MSRDataset} constructed a multi-aspect developer sentiment dataset based on \textsc{JIRA} comments and sentences, which is regarded as one of the golden dataset\cite{gold} for sentiment analysis of the software engineering domain. The authors also analyzed the impact of sentiments. For example, they found impolite comments\cite{Ortu15Politeness,politepeerj} and bullies\cite{Ortu15MSRBullies} are related to longer issue fixing time. Meanwhile, certain combinations of VAD\cite{Ortu16MSRVAD} (Valence-Arousal-Dominance) scores may indicate longer issue resolution time, as well as productivity problems such as burnout. Such results in line with Khan \textit{et al.}'s research \cite{KhanMood} reporting developers performs better with higher degree of arousal and valence after doing physical exercises. In further research on \textsc{GitHub} comments\cite{Ortu18UsersCommenters}, the authors also managed to differentiate comments made by users and developers using their emotions, sentiments and degree of politeness. 
\begin{figure*}[htbp]
\centerline{\includegraphics[height=6cm ,width=14cm,angle=0]{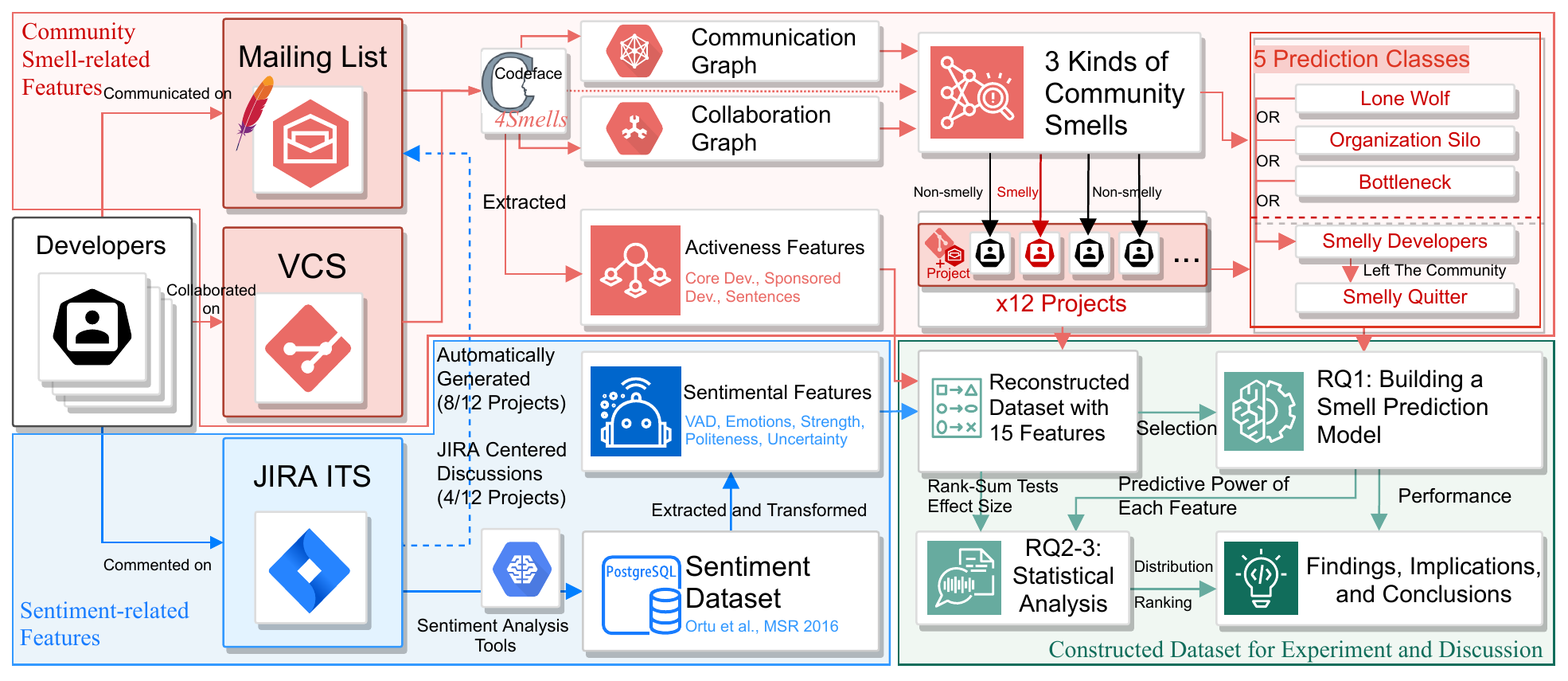}}
\caption{Dataset construction and its usage.}
\label{fig}
\end{figure*}

Happiness is, in most cases, related to positive outcomes of software engineering\cite{graziotin2014happy,Graziotin_17_SEMOTION}. However, too many positive comments may also lead to buggy code\cite{Huq}. Graziotin \textit{et al.}\cite{Graziotin_17_SEMOTION, GraziotinJSS18} identified 49 consequences of unhappiness in their qualitative research. Internal consequences include low motivation, low cognitive performance, and work withdrawal. External ones contain low code quality and code discharging, \textit{i.e.}, destroying the codebase. Studies also showed that negative emotions are signals of the appearance of bug fix-inducing changes, but they may result in a longer issue-fixing time\cite{Huq} and issue-reopening\cite{Cheruvelil19}. As for negative sentimental expressions in code refactoring, Singh \textit{et al.}\cite{Singh_APSEC17_TOOL_NEGATIVE_SENTIMENT} concluded that poor tooling, error-proneness and high complexity of such activities could cause negative sentiments. Controversially, Islam \textit{et al.} \cite{SERA_16} reported opposite observations. Additionally, they discovered that the sentiment of code commit varies from commit types, \textit{e.g.}, there exist higher negative emotions for new feature implementation. 

To sum up, it is possible to detect developers' sentiments, and sentiments do have affections on the software development process. However, a generally-accepted theorem explaining the pattern of such affection has not been proposed at present. 

\section{Dataset Construction}

This section describes how we export and process the developers' sentiment features and community smells' occurrences to define and train our machine learner in the next section. Fig. 1 depicts generally the origins and construction methods of our dataset, as well as its further usage in this paper.

\subsection{Extracting Developer Sentiment Features}\label{AA}


First, we recover and adjust the raw data, \textit{i.e.}, sentiments of every developer of specific projects in \textsc{JIRA} ITS located in 2 tables called \texttt{jira\_issue\_comment} and \texttt{jira\_user}.  Afterwards, we also group the features by developers and projects to prepare for cross- and within-project prediction. Details of the features will be described in Section IV. 

\subsection{Selecting Projects and Fetching Mailing Lists}

\begin{table*}[htbp]
  \centering
  \caption{The Projects Analyzed in this Paper}
    \begin{tabular}{c|c|c|c|c|c}
    
    \hline
    \multicolumn{1}{l|}{\textbf{Repository}} & \textbf{Project Name} & \multicolumn{1}{l|}{\textbf{Developers}} & \textbf{Date Range} & \multicolumn{1}{l|}{\textbf{Issue Reports}} & 
    \multicolumn{1}{c}{\textbf{  Mailing List Name}}  \\

    \hline
    \textsc{ASF}   & HBase & 919   & 2007-04 - 2014-04 & 68806 & hbase-dev\footnotemark[1]\\
    \textsc{ASF}   & Hadoop Common & 1221  & 2009-05 - 2014-02 & 61905 & commons-dev\footnotemark[1]\\
    \textsc{ASF}   & Hadoop HDFS & 745   & 2009-05 - 2011-04 & 42188 & hadoop-hdfs-dev\footnotemark[1]\\
    \textsc{ASF}   & Cassandra & 1161  & 2009-03 - 2014-03 & 41937 & cassandra-dev\footnotemark[1]\\
    \textsc{ASF}   & Hadoop Map/Reduce & 857   & 2009-05 - 2011-03 & 34747 & hadoop-mapreduce-dev\footnotemark[1]\\
    \textsc{ASF}   & Hive  & 839   & 2008-09 - 2014-03 & 34449 & hive-dev\footnotemark[1]\\
    \textsc{ASF}   & Harmony & 306   & 2005-09 - 2011-07 & 28325 & harmony-dev\footnotemark[1]\\
    \textsc{ASF}   & OFBiz & 538   & 2006-07 - 2014-02 & 25667 & ofbiz-dev\footnotemark[1]\\
    \textsc{JBoss} & Hibernate ORM & 3958  & 2007-06 - 2014-01 & 23549 & hibernate-dev\footnotemark[2]\\
    \textsc{ASF}   & Camel & 882   & 2007-04 - 2014-01 & 21758 & camel-dev\footnotemark[1]\\
    \textsc{ASF}   & Wicket & 1210  & 2006-10 - 2014-01 & 17030 & wicket-dev\footnotemark[1]\\
    \textsc{ASF}   & Zookeeper & 484   & 2005-09 - 2011-07 & 13634 & zookeeper-dev\footnotemark[1]\\
    \hline
    
    \end{tabular}%
    
  \label{tab:addlabel}%
\end{table*}%

The sentiment dataset consists of project comments in ITS of 4 major open source foundations namely \textsc{ASF}, \textsc{Codehaus}, \textsc{JBoss}, and \textsc{Spring}. However, only 23 projects contain sentiment data. Among the 23 projects, we first exclude 3 projects whose mailing list service providers do not support archive extraction. Next, we exclude 4 more projects as their mailing lists do not cover the time range of their sentiment data. Afterwards, we remove 3 software projects as they were sharing the same instances of \textsc{JIRA} ITS or Version Control System (VCS) repository with other projects, and such projects were incompatible with \textsc{Codeface4Smells}. Finally, We also filter out 1 project whose VAD data is missing in the dataset.

To sum up, the actual 12 projects we use to perform the analysis are listed in Table I.  We fetch their mailing lists from the archives provided by their open source foundations. 

\subsection{Detecting community smells}
Although we are aware of the existence of other community smell detection tools or models \cite{ICGSE,KBS}, they are not publicly available. Thus, we apply the state-of-the-art tool \textsc{Codeface4Smells}\cite{PalombaCSDetect,PalombaMasters} to detect community smells.

We follow strictly the instructions of the \textsc{Codeface4Smells} repository, \textit{e.g.}, we execute the application in the suggested \textsc{Vagrant} instance, and we fix the broken dependencies in order to avoid platform-specific problems. The only modification is adding exportation features to the socio-technical analysis script to derive names and e-mails of developers affected by each community smell. 

As for configurations, community smell analysis must be performed in a given window. In our case, the window is 3 months as previous works suggested\cite{PalombaCSDetect,PalombaMasters,PalombaCSPredict}. Similar as the tool's replication package did \cite{PalombaMasters}, we also specify every commit to analyze in configuration files, the commits are exported from \textsc{Git} repositories of the projects using \textsc{PyDriller} \cite{pydriller}, which is also a commonly applied tool in software repository mining tasks\cite{PalombaMSR}. The configuration files are also provided in our replication package\cite{replication}.

\textsc{Codeface4Smells} is able to detect 5 community smells, including \textit{Organization Silo}, \textit{Lone Wolf}, \textit{Bottleneck}, \textit{Black Cloud}, and \textit{Prima Donnas}. However, \textit{Prima Donnas} detection is not empirically\cite{PalombaCSDetect} proved effective since its first appearance\cite{PalombaMasters}, so we do not take this community smell into consideration. \textit{Black Cloud} is sparsely distributed in software systems\cite{PalombaCSDetect}. In our dataset, there are several \textit{Black Cloud} appearances. However, the developer affected is not captured in the sentiment dataset, thus we could not perform \textit{Black Cloud} prediction. Consequently, our research scope includes 3 community smells, namely \textit{Organizational Silo}, \textit{Lone Wolf}, and \textit{Bottleneck}. We also involve \textit{Smelly Developers} and \textit{Smelly Quitters} as additional prediction classes, the details will be described in section IV.

\section{Design and Evaluation of a Developer-Oriented and  Sentiment-aware Community Smell Prediction Model}
The goal of our study is to evaluate to what extent the occurrence of community smells on developers can be predicted by their sentiments, with the purpose of understanding the impact of sentiment to community smells from the granularity and perspective of developer themselves, in the mean time, provide researchers and practitioners with novel aspects and suggestions on community healthiness. This section proposes our research questions and our methodology.
  \footnotetext[1]{ http://mail-archives.apache.org/mod\_mbox/\{Mailing List Name\}}
  
  \footnotetext[2]{ https://lists.jboss.org/archives/list/hibernate-dev@lists.jboss.org/}

\subsection{Research Questions and Methodologies}

\begin{framed}
 RQ1: To what extent can we predict the occurrence of community smells on developers using their sentiments?
\end{framed}

Starting from the dataset we build in Section III, we define dependent and independent variables, and we build prediction models using Machine Learning classifiers. To pick the most appropriate machine learner, and to figure out why our model works or fails, their performances must be assessed. Hence, we propose RQ2. 

\begin{framed}
RQ2: What is the predictive power of sentiments to the occurrence of community smells on developers?
\end{framed}

We investigate further the predictive power of each feature to reveal the contribution of every sentimental feature to our prediction model. The conclusion may provide us with useful information to explain our model and to look deeply into the distribution of sentimental features, which leads us to RQ3.  

\begin{framed}
RQ3: Are smelly and non-smelly developers different in terms of their sentiments?
\end{framed}

Apart from model performance metrics, we intend to explain how sentiments impact the healthiness of development community according to their statistical characteristics including data distributions and mean values, and to differentiate smelly and non-smelly developers. Furthermore, we attempt to make suggestions to practitioners based on our findings.

\begin{table*}[htbp]
  \centering
  \caption{Features Extracted from the Developer Sentiment dataset and \textsc{\textsc{Codeface4Smells}}-defined Metrics}
    \begin{tabular}{|l|c|l|}
    \hline
        \multicolumn{3}{|l|}{\textbf{VAD.} Automatically detected, the summation of mean value of sentences in comments by extending lexicons\cite{Ortu16MSRVAD} in \textsc{SentiStrength}  \cite{SentiStrengthOrig}.} \\
    \hline
    Mean Valence   & VAL & The mean intensity of valence, \textit{i.e.}, how developers enjoy a situation. \\
    Mean Arousal   & ARO & The mean intensity of arousal, \textit{i.e.}, increased alertness. \\
    Mean Dominance   & DOM & The mean intensity of dominance, the extent that developers were feeling in control. \\
    \hline
    \multicolumn{3}{|l|}{\textbf{Emotions.} Binary value labeled manually in the dataset.} \\
    \hline
    Mean Sadness & SAD   & The mean intensity of all sadness expressions. \\
    Mean Anger & ANG   & The mean intensity of all angry expressions. \\
    Mean Love & LOV   & The mean intensity of all love expressions. \\
    Mean Joy & JOY   & The mean intensity of all joy expressions. \\
    \hline
    \multicolumn{3}{|l|}{\textbf{Sentiment Strength.} Measured using \textsc{SentiStrength}\cite{SentiStrengthOrig}, ranges between [-1;1].} \\
    \hline
    Mean Postive Sentiment & POS   & The mean intensity of all sentiments smaller than 0. \\
    Mean Negative Sentiment & NEG   & The mean intensity of all sentiments greater than 0. \\
    \hline
    \multicolumn{3}{|l|}{\textbf{Politeness.} Binary value measured using Danescu \textit{et al.}’s tool \cite{politeness}.} \\
    \hline
    Politeness Proportion & POL   & The proportion of polite expressions in all commentary sentences of a developer. \\
    \hline
    \multicolumn{3}{|l|}{\textbf{Uncertainty.} 4 categorical Grammatical Moods (GM) and modality in [-1;1]  measured using \textsc{Pattern} \cite{mood} by detecting auxiliary verbs and adverbs.} \\
    \hline
    Indicative GM Proportion & IND   & The proportion of sentences that express fact or belief, \textit{e.g.}, It rains. \\
    Imperative GM Proportion & IMP   & The proportion of sentences that express command, warning, \textit{e.g.}, Do not rain! \\
    Conditional GM Proportion & CON   & The proportion of sentences in the form like would, may, or will, \textit{e.g.}, It might rain. \\
    Subjunctive GM Proportion & SUB   & The proportion of sentences in the form like wish, were, \textit{e.g.}, I hope it rains. \\
    Mean Degree of Modality & MOD   & The degree of uncertainty of a sentence, ranges between [-1;1]. \\
    \hline
    \multicolumn{3}{|l|}{\textbf{Activeness.} Control variables reflecting developers' working state other than sentiments.} \\
    \hline
    Total Sentences & SEN   & Number of total senteces developers commented in the \textsc{JIRA} ITS, extracted from \cite{Ortu16MSRDataset}. \\
    Core Ranges Count & COR   & Number of total ranges developer acted as a core developer, measured by \textsc{Codeface4Smells}\cite{PalombaCSDetect, PalombaMasters}. \\
    Sponsored Ranges Count & SPO   & Number of total ranges developer commited only in working hours\cite{PalombaCSPredict}, measured by \textsc{Codeface4Smells}. \\
    \hline
    \end{tabular}%
  \label{tab:addlabel}%
\end{table*}%

\subsection{RQ1: Defining and Evaluating the Proposed Model in Cross- and Within-project Scenarios}

\subsubsection{Dependent Variables} In this paragraph, we list the definition of 5 smell-related prediction classes of our model. The 3 concerned community smells are defined as follows:
\begin{itemize}
\item \textbf{Organizational Silo}: refers to the
presence of siloed areas of the developer community
that do not communicate, except through one or two
of their respective members\cite{PalombaCSDetect}, \textit{i.e.}, co-committing developers do not directly communicate at all\cite{PalombaMasters}.
\item \textbf{Lone Wolf}: reflects co-committing software developers
who exhibit uncooperative behaviour and mistrust by not
appropriately communicating\cite{PalombaCSDetect}, \textit{i.e.}, the collaboration edges that do not have a communication counterpart\cite{PalombaMasters}.
\item \textbf{Bottleneck}: unique boundary spanners interpose
themselves into every interaction across sub-communities\cite{PalombaCSDetect}.
\end{itemize}

Despite the 3 smells, we also export 2 related classes: 

\begin{itemize}
\item \textbf{Smelly Developer}: developers affected by any of the 3 above-mentioned community smells.
\item \textbf{Smelly Quitter}: \textit{Smelly Developers} from the last analysis window who quit the community in current window\cite{PalombaCSPredict}.
\end{itemize}
\begin{figure}[t]
\centerline{\includegraphics[height=3.5cm ,width=6.02cm,angle=0]{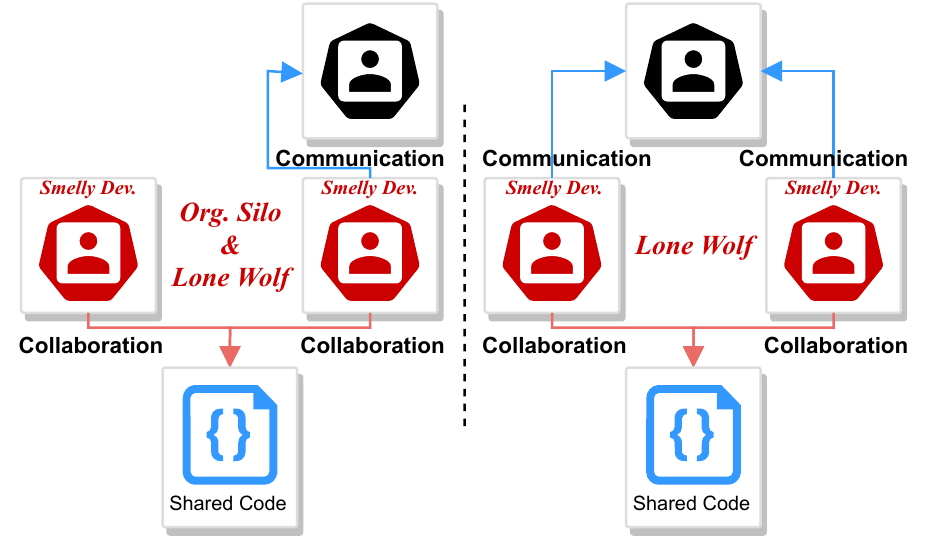}}
\caption{Lone Wolf and Organizational Silo identification pattern.}
\label{fig}
\end{figure}

Fig. 2 illustrates an example of community smell detection, in which smelly developers are marked in red. The left part of Fig. 2 shows a siloed area of developers, while the right part illustrates 2 lone wolves collaborating on shared code with indirect communication. To clarify, \textit{Organizational Silo} is a subset of \textit{Lone Wolf}\cite{PalombaMasters}. Technically, the major difference between the two smells stands in the definition of lacking  connectivity in the communication graph. \textit{Organizational Silo} is detected if collaborating developers are disconnected in the  communication graph. \textit{Lone Wolf} is detected if collaborating developers are not neighbours in the communication graph, \textit{i.e.}, lacks 1-degree connection.  Differently, \textit{Bottleneck} is detected based on completely the communication graph, and its definition is remarkably similar to the unique
boundary spanner in social-network analysis \cite{PalombaCSDetect}.

Our motivation to involve \textit{Smelly Quitter} prediction is that literature reported negative sentiments may cause developer to ``destroy codebase'' \cite{Graziotin_17_SEMOTION, GraziotinJSS18}, and to ``quit over mistreatment'' \cite{Graziotin_14_QUIT}. Similarly, we assume it may also lead to the discharge of community. 

\subsubsection{Independent Variables} We integrate sentimental and activeness features in Table II as independent variables. 

Similar to \cite{SERA_16}, we split positive and negative sentiment into 2 features, POS and NEG. 

We do not include an impoliteness proportion feature because the politeness labels available in the dataset are either polite or impolite, which will result in high correlation if we involve both. We ignore the confidential of politeness score since Ortu \textit{et al.} \cite{Ortu16MSRDataset} have already dropped the results with confidential less than a conventional threshold (0.5)\cite{Ortu15Politeness}.

The dataset also contains a mood column whose proper meaning was not clearly described in the original paper\cite{Ortu15MSRBullies}. We look into the documentation of the detection tool\cite{mood} and confirm this feature measures the GM of a sentence, and the result is mapped into 4 classes, \textit{i.e.}, \{indicative, imperative, conditional, subjunctive\}. In the dataset the 4 classes are presented in 4 values, namely \{0,1,2,3\}. We map the mood attribute into 4 proportional features to measure the developers' characteristics of expression. 

Finally, we also introduce 3 activeness features including total sentences of developers commented on ITS (SEN) \cite{Ortu16MSRDataset}, the count of ranges that developers acted as core (COR) and sponsored (SPO) developers\cite{PalombaCSDetect,PalombaCSPredict}. They are used as control variables to assess the actual impact of sentimental features in comparison with activeness.

\subsubsection{Data Balancing and Feature Selection} Our dataset is highly imbalanced, which in line with the observations in related researches \cite{KBS,PalombaCSPredict}. Smelly developers account for 4.80\% of the overall developer population, which may hinder model performance. Therefore, we preprocess our data with Random Undersampling strategy, which is proved beneficial to imbalanced data in software engineering\cite{imbalanced}. We also address potential multicollinearity problem by removing the correlated features as previous researches proposed\cite{PalombaCSPredict}.

\subsubsection{Scenarios} We build models separately in both cross-project and within-project scenarios. For cross-project tasks, we merge the developers' features regardless of the projects they are working on. Notably, the same developers appearing on multiple projects are treated differently. 

\subsubsection{Training Machine Learners} 
Previous researches in community smell and code smell detection showed Random Forest was the best-performed classifier\cite{PalombaMSR,PalombaCSPredict}. The reason we evaluate again the different classifiers is that our data derive from developers individually, which is different from previous scenarios. Moreover, predictors have different characteristics in terms of design and effectiveness, \textit{i.e.}, the risk of overfitting and different execution speed\cite{PalombaCSPredict}. We intend to discover the best classifier for our scenario.

We apply the \textsc{scikit-learn} package\cite{scikit} from \textsc{Python} to train machine learners using multiple classifiers that have been used in previous works\cite{PalombaMSR,PalombaCSPredict,smellbugprediction}, including Random Forest, Decision Tree, Support Vector Machine, Multilayer Perceptron, Adaboost, Naive-Bayes, and Logistic Regression.

As related work suggested\cite{PalombaMSR}, instead of using default settings, we configure the hyper-parameters of the classifiers by exploiting Exhaustive Grid Search with a 10-Fold Cross-Validation strategy to calculate multiple times the performance of every combination of parameters. 
\subsubsection{Validation and Assessment} To validate the performance of each classifier, we employ a 10 $\times$ 10-Fold Cross-Validation strategy to make sure that we have unbiased and stable \cite{PalombaCSPredict,esd1} performance data in cross-project scenarios. For the within-project scenario, we apply the Leave-One-Out Cross-Validation (LOOCV) according to \cite{PalombaCSPredict,esd1}  which is proved to be stable and the least biased one in this scenario. Finally, we compute classical performance metrics including precision, recall, F-Measure, and AUC-ROC\cite{aucroc} to pick the best-ranked classifier in both scenarios. Thus, we could answer RQ1 quantificationally.

\subsection{RQ2: Explaining the Predictive Power of Features}
To answer this RQ, we need to explain the extent of predictive power that each independent variable contribute to the best-performed classifier in RQ1. 

First, we apply an information gain algorithm, which has already been used in previous researches of community smells\cite{PalombaCSPredict} and code smells\cite{PalombaMSR} to compute the gain provided by each feature within the model to the correct prediction of a dependent variable. The approach measures the amount of reduction in uncertainty, \textit{i.e.}, Shannon's Entropy, of the model concerned after involving in a new feature. This paper applies the mutual information classification \cite{MI} implementation of \textsc{scikit-learn} package, which is originally designed for feature selection and identical in definition to the Gain Ratio Feature Evaluation used by previous researches\cite{PalombaCSPredict,PalombaMSR}. The algorithm ranks the features in descendent order according to each of their contribution to the entropy reduction of the concerned model. Thus, the features contributing the most predictive power could be identified.  

Then, in order to assess the significance of the ranking of information gain, we also involve the Scott-Knott Effect Size Difference (SK-ESD) test. Scott-Knott test \cite{SK} is a statistical measure to compare and differentiate model performance using a hierarchical clustering approach to group the means of assessment metrics, \textit{e.g.}, F-Measures of multiple models. Scott-Knott test assumes input data distribution to be normally distributed, thus may be ineffective for non-normally distributed data. The ESD test is an enhanced version of the original Scott-Knott test that corrects the non-normal distribution of the input to make it comply with the requirements executing the Scott-Knott test. Meanwhile, it uses Cliff's Delta\cite{cliff} as an effect size measure to merge groups having negligible effect sizes. Scott-Knott and SK-ESD tests have already been applied in various researches in code smell \cite{PalombaMSR}, community smell \cite{PalombaCSPredict}, and software defect \cite{YangGe}. We use the original implementation of Tantithamthavorn \textit{et al.}\cite{esd1,esd2} available on \textsc{GitHub}. 

Both information gain and SK-ESD algorithms are executed on each prediction class independently. We report the results from both cross- and within-project scenarios. 

\subsection{RQ3: Investigating the Difference between Smelly and Non-Smelly Developers in Terms of Sentiments}

\begin{figure}[t]
\centerline{\includegraphics[height=7.53cm ,width=9cm,angle=0]{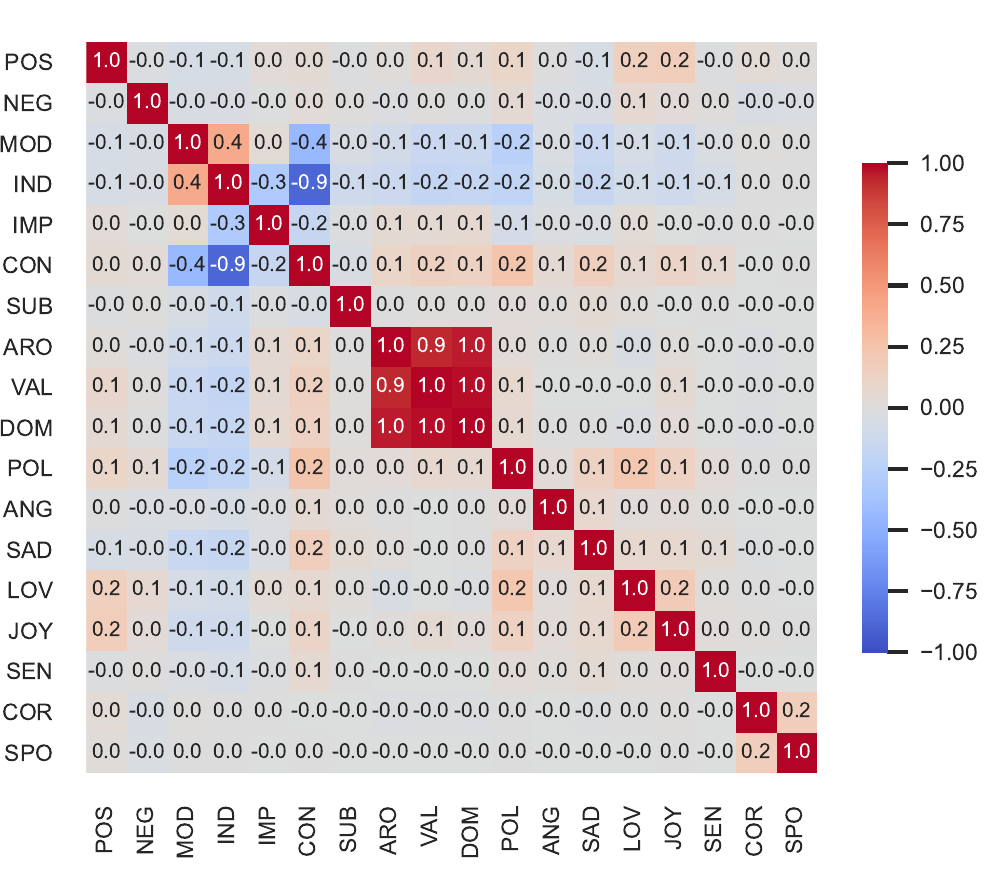}}
\caption{Correlation heat-map of features.}
\label{fig}
\end{figure}

\begin{figure*}[htbp]
\centerline{\includegraphics[height=6.5cm ,width=13cm,angle=0]{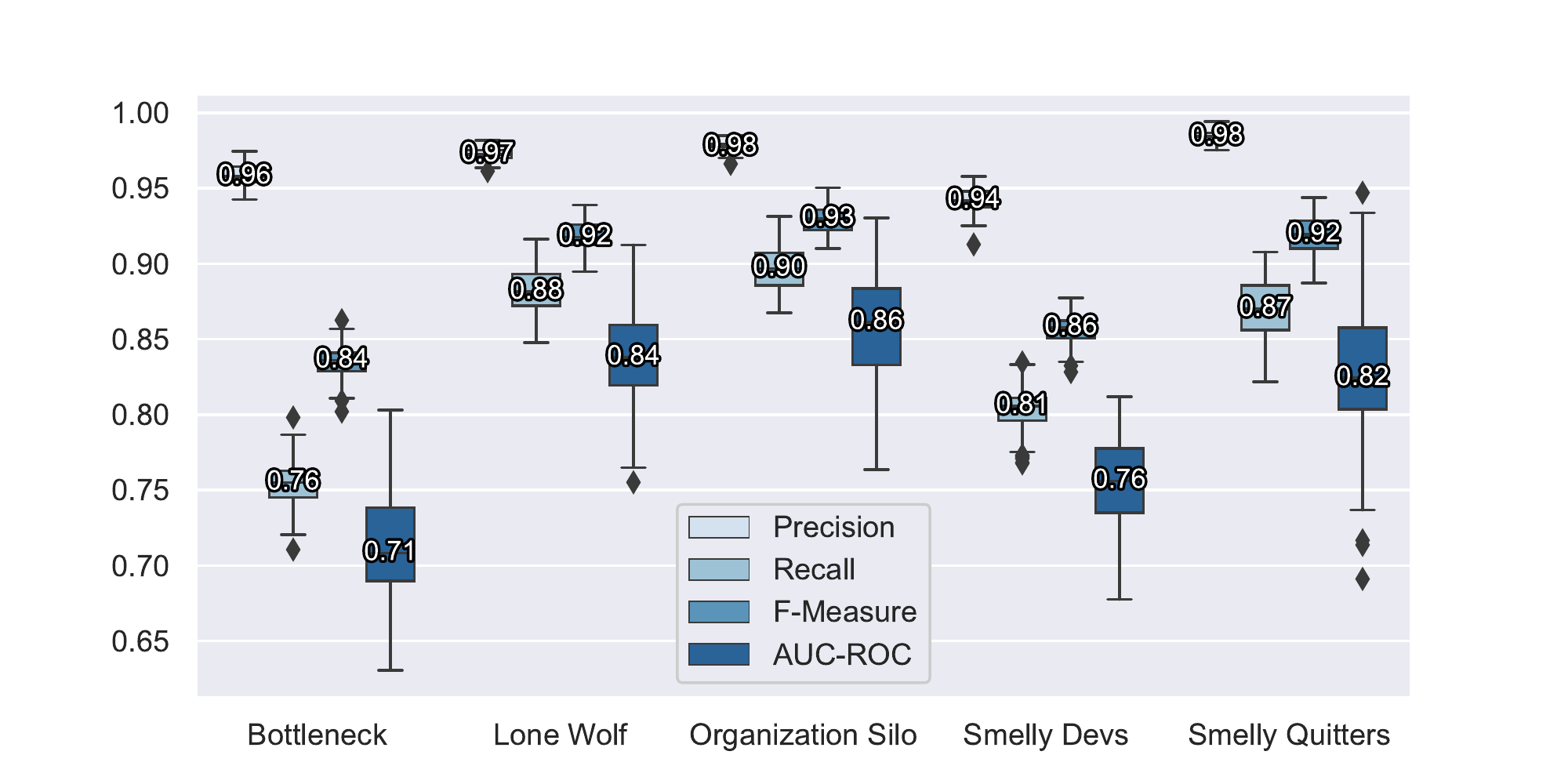}}
\caption{Cross-project prediction performance.}
\label{fig}
\end{figure*}

\begin{figure*}[htbp]
\centerline{\includegraphics[height=6.5cm ,width=13cm,angle=0]{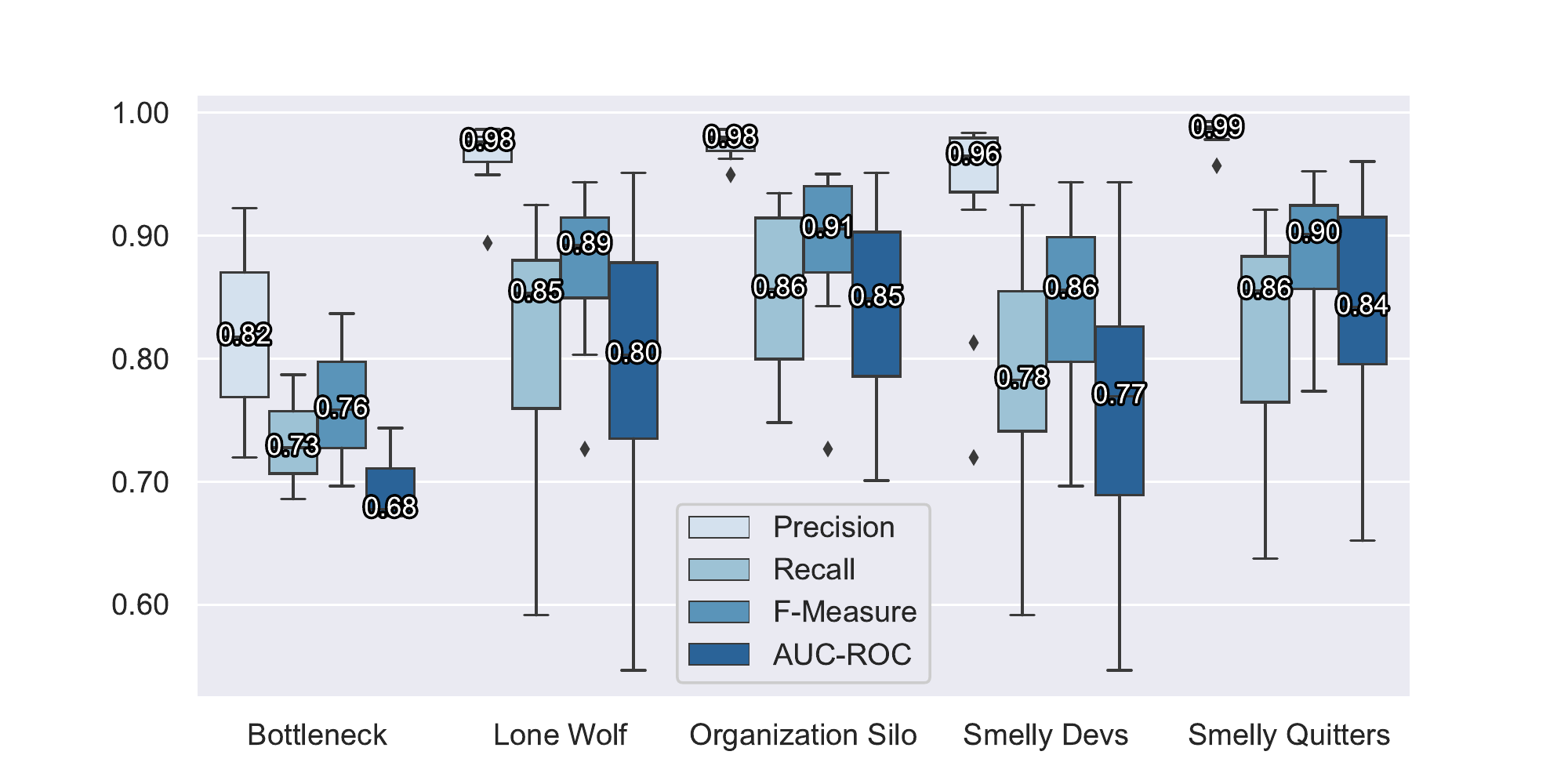}}
\caption{Within-project prediction performance.}
\label{fig}
\end{figure*}
We take a closer look at the sentimental features' difference in distribution and intensity for smelly and non-smelly developers. We apply statistical measures, \textit{i.e.}, Wilcoxon Ranksum Test ($\alpha$ = 0.05, p $\textless$ $\alpha$ for statically significant \cite{scent}) to analyze the significance of the difference in distribution. Meanwhile, we calculate Cliff's Delta ($\delta$) for each test to measure the effect size, \textit{i.e.}, the extent of the difference, which is negligible when $\mid$$\delta$$\mid$ $\textless$ 0.147, small when 0.147 $\leq$ $\mid$$\delta$$\mid$  $\textless$ 0.33, medium when 0.33 $\leq$ $\mid$$\delta$$\mid$  $\textless$ 0.474, and large otherwise. Such measures have been applied in previous works of developer sentiment \cite{Ortu18UsersCommenters,Ortu16MSRVAD} and code smell\cite{scent}. We also report the mean and variance values of the features.

We expect to find statistical significance from our dataset to support our findings in RQ2, and to provide practitioners with advice to prevent community smells from occurring at an early stage.

\section{Result and Discussion}
In this section, we demonstrate and discuss the results of our experiment, answer the proposed research questions, and describe our findings. 
\subsection{RQ1: Model Performance}

 In Fig. 3, we assess the correlations of our features. Result shows the VAD features, \textit{i.e.}, VAL, ARO, DOM, correlate with each other ($\rho$\textgreater0.9).  Although we find removing any 2 of them does not cause a significant change in the classifier's performance (\textless 2\%), we still exclude VAL and DOM to avoid potential multicollinearity problems. We also exclude CON for the same reason. Consequently, we remove 3 features out of 18 original ones.

We train multiple classifiers on the dataset and report the result of the best performed one, \textit{i.e.}, Random Forest, in Fig. 4 and Fig. 5 for both cross- and within-project prediction, and we include results for the 2 scenarios in the sets of box-plots. We also present the median of weighted average performance of our models in Table III.

Due to inconsistent community smell appearance in different projects, the performance of within-project prediction is less stable than cross-project ones, revealing potential drawbacks of our approach such as the capability to deal with cold start problem, \textit{i.e.,} do not have sufficient useful data to start with, which should be addressed in future work. 

However, the median values of the model's performances show that the model still works in most cases.

\begin{framed}
Finding 1. Our prediction model could predict the occurrence of community smells on developers in most cases. In terms of cross-project prediction, our model reaches mean F-Measures ranging from 84\% to 93\%. Meanwhile, the model achieves F-Measures from 76\% to 91\% for within-project prediction. 
\end{framed}

\subsection{RQ2: Features' Predictive Power}

\begin{table}[!b]
  \centering
  \caption{Weighted Average of Model Performance}
    \begin{tabular}{c|c|c|c|c|c}
    \hline
    \textbf{Class} & \textbf{Scenario} & \textbf{Prec.} & \textbf{Rec.} & \textbf{F-Meas.} & \textbf{AUC-ROC} \\
    \hline
        \multirow{2}{*}{Bottleneck} & Within & .96    & .76    & .84    & .71 \\
          & Cross & .82    & .73    & .76    & .68 \\
    \hline
    \multirow{2}{*}{Lone Wolf} & Within & .97    & .88    & .92    & .84 \\
          & Cross & .98    & .85    & .89    & .80 \\
    \hline
      \multirow{2}{*}{Org. Silo} & Within & .98    & .90    & .93    & .86 \\
          & Cross & .98    & .86    & .91    & .85 \\
    \hline
    \multirow{2}{*}{Smelly Dev.} & Within & .94    & .81    & .86    & .76 \\
          & Cross & .96    & .78    & .86    & .77 \\
    \hline
    \multirow{2}{*}{Smelly Quitter} & Within & .98    & .87    & .92    & .82 \\
          & Cross & .99    & .86    & .90    & .84 \\
    \hline
    \end{tabular}%
  \label{tab:addlabel}%
\end{table}%
\begin{table}[!b]
  \centering
  \caption{Ranking and Mean Gain Ratio of Each Feature}
    \begin{tabular}{c|c|c|c||cccc}
    \hline
    \textbf{Metric} & \textbf{Mean} & \textbf{W.-P.} & \textbf{C.-P.} & \multicolumn{1}{c|}{\textbf{Metric}} & \multicolumn{1}{c|}{\textbf{Mean}} & \multicolumn{1}{c|}{\textbf{W.-P.}} & \textbf{C.-P.} \\
    \hline
    \textbf{\textit{IMP}} & .18  & 1     & 1     & \multicolumn{1}{c|}{\textit{\textbf{ANG}}} & \multicolumn{1}{c|}{.12} & \multicolumn{1}{c|}{4} & 4 \\
    \textbf{\textit{SAD}} & .18  & 1     & 1     & \multicolumn{1}{c|}{\textit{\textbf{POS}}} & \multicolumn{1}{c|}{.11} & \multicolumn{1}{c|}{5} & 5 \\
    \textbf{\textit{POL}} & .18  & 1     & 1     & \multicolumn{1}{c|}{\textit{\textbf{MOD}}} & \multicolumn{1}{c|}{.08} & \multicolumn{1}{c|}{6} & 6 \\
    \textbf{\textit{IND}} & .18  & 1     & 1     & \multicolumn{1}{c|}{\textit{\textbf{SUB}}} & \multicolumn{1}{c|}{.07} & \multicolumn{1}{c|}{7} & 7 \\
    \textbf{\textit{LOV}} & .18  & 1     & 2     & \multicolumn{1}{c|}{\textit{\textbf{ARO}}} & \multicolumn{1}{c|}{.03} & \multicolumn{1}{c|}{8} & 8 \\
    \textbf{\textit{JOY}} & .17  & 2     & 1     & \multicolumn{1}{c|}{COR} & \multicolumn{1}{c|}{.03} & \multicolumn{1}{c|}{9} & 9 \\
    \textbf{\textit{NEG}} & .13  & 3     & 2     & \multicolumn{1}{c|}{SPO} & \multicolumn{1}{c|}{.00} & \multicolumn{1}{c|}{10} & 10 \\
    SEN   & .13  & 3     & 3     & \multicolumn{1}{c|}{} & \multicolumn{1}{c|}{} & \multicolumn{1}{c|}{} &  \\
    \hline
    \end{tabular}%
  \label{tab:addlabel}%
\end{table}%

Table IV lists the information gains that features contribute to our prediction model in descendent order. Metrics in bold and italic are \textbf{\textit{Sentimental Features}}. Column W.-P. \& C.-P. list the SK-ESD rank of gain of features to the within- and cross-project model. Fig. 6 and Fig. 7 depict the mean (in solid lines), median (in dashed lines), as well as the rank of each feature's gain. Metrics given the same rank are marked in the same color. The ranks of information gains in different classes are similar, so we do not report them due to the limitation of space. Therefore, the gains presented in this section is to describe the contribution of metrics to the overall performance of our model.

Every sentimental feature has some contribution to the model (mean Gain Ratio $\geq$ 0.033), and their contributions are average and moderate. Meanwhile, the rankings are slightly different in the 2 scenarios. However, 6 metrics, \textit{i.e.}, IMP, SAD, POL, IND, LOV, JOY, are always stronger predictors than any of the activeness features. Meanwhile, they are also among the highest ones. 

In particular, imperative GM is the strongest factor among all the features in terms of predictive power, while indicative GM is in the same rank, showing certainty is a determining aspect of community healthiness. 

Either positive or negative emotions, including sadness, love, and joy, are related to community smell issues. 

POL came after IMP and SAD, which is in line with researches \cite{politeness,politepeerj,Ortu15Politeness,Ortu15MSRBullies} suggesting politeness may have an impact on developers' productivity. 

The above-mentioned sentimental features contribute more than the activeness ones, \textit{i.e.}, SEN, COR and SPO, indicating the centrality and discussion activeness of developers are weaker predictors than several sentiments. Recent work \cite{ahmed} also reported that the number of commits are related to community smells' occurrence. In response, we are planning to investigate the traditional process metrics' impact on developers' working state in the software community. Since COR measures the centrality of developers in the communication and the collaboration graphs, we assume it is likely to be correlated with commit activeness over software systems. 
\begin{figure}[!b]
\centerline{\includegraphics[height=3.3cm ,width=10cm,angle=0]{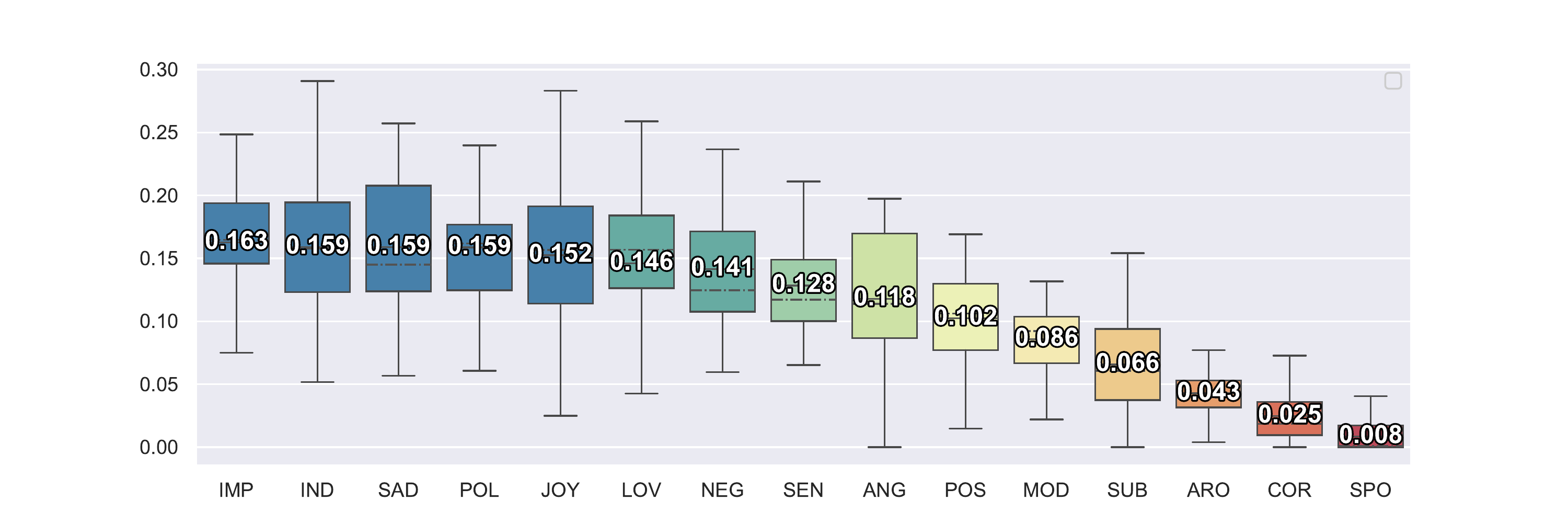}}
\caption{Within-project Information gain classified by SK-ESD.}
\label{fig}
\end{figure}

\begin{figure}[!b]
\centerline{\includegraphics[height=3.3cm ,width=10cm,angle=0]{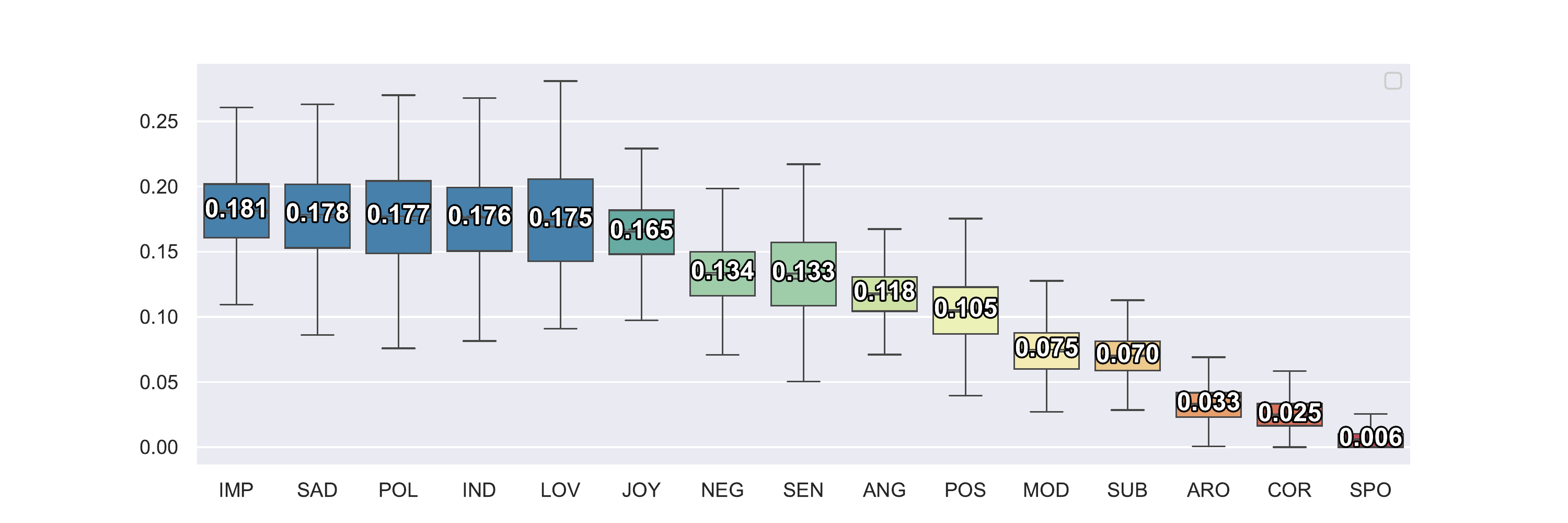}}
\caption{Cross-project Information gain classified by SK-ESD.}
\label{fig}
\end{figure}
\begin{framed}
Finding 2. In terms of community smells' occurrence on developers, 6 sentimental features are stronger predictors than activeness ones. Imperative and indicative GM are among the strongest predictors, indicating certainty a key factor to community healthiness. Politeness and both positive and negative emotions including sadness, love, and joy, are also powerful predictors. 
\end{framed}

\begin{table*}[htbp]
  \centering
  \caption{Result of Statistical Analysis}
    \begin{tabular}{l|r|r|r|r|r|r|r|r|r|r|r|r}
    \hline
    \textbf{Features} & \multicolumn{2}{c|}{\textbf{IMP}} & \multicolumn{2}{c|}{\textbf{POL}} & \multicolumn{2}{c|}{\textbf{LOV}} & \multicolumn{2}{c|}{\textbf{JOY}} & \multicolumn{2}{c|}{\textbf{NEG}} & \multicolumn{2}{c}{\textbf{IND}} \\
    \hline
    Wilcoxon R. & \multicolumn{2}{c|}{p\textless0.001} & \multicolumn{2}{c|}{p\textless0.001} & \multicolumn{2}{c|}{p\textless0.001} & \multicolumn{2}{c|}{p\textless0.001} & \multicolumn{2}{c|}{p\textless0.001} & \multicolumn{2}{c}{p\textless0.001} \\
    \hline
    Cliff's Delta & -0.48 & \multicolumn{1}{c|}{L} & -0.25 & \multicolumn{1}{c|}{S} & 0.34  & \multicolumn{1}{c|}{M} & 0.38  & \multicolumn{1}{c|}{M} & -0.09 & \multicolumn{1}{c|}{-} & -0.27 & \multicolumn{1}{c}{S} \\
    \hline
    
  \textbf{Smelly Dev.?} & \multicolumn{1}{c|}{\textbf{Y}} & \multicolumn{1}{c|}{\textbf{N}} &
\multicolumn{1}{c|}{\textbf{Y}} & \multicolumn{1}{c|}{\textbf{N}} &\multicolumn{1}{c|}{\textbf{Y}} & \multicolumn{1}{c|}{\textbf{N}} &\multicolumn{1}{c|}{\textbf{Y}} & \multicolumn{1}{c|}{\textbf{N}} &\multicolumn{1}{c|}{\textbf{Y}} & \multicolumn{1}{c|}{\textbf{N}} &\multicolumn{1}{c|}{\textbf{Y}} & \multicolumn{1}{c}{\textbf{N}} \\
    \hline
    Mean  & 0.09  & 0.26  & 0.54  & 0.67  & 0.14  & 0.13  & 0.08  & 0.02  & -0.16 & -0.17 & 0.65  & 0.77 \\
    \hline
    Variance & 0.01  & 0.08  & 0.04  & 0.08  & 0.04  & 0.09  & 0.06  & 0.03  & 0.01  & 0.02  & 0.33  & 0.06 \\
    \hline    \hline

    \textbf{Features} & \multicolumn{2}{c|}{\textbf{SAD}} & \multicolumn{2}{c|}{\textbf{ANG}} & \multicolumn{2}{c|}{\textbf{POS}} & \multicolumn{2}{c|}{\textbf{MOD}} & \multicolumn{2}{c|}{\textbf{SUB}} & \multicolumn{2}{c}{\textbf{ARO}} \\
    \hline
    Wilcoxon R. & \multicolumn{2}{c|}{p\textless0.001} & \multicolumn{2}{c|}{p\textless0.001} & \multicolumn{2}{c|}{p\textless0.001} & \multicolumn{2}{c|}{p\textgreater0.05} & \multicolumn{2}{c|}{p\textless0.001} & \multicolumn{2}{c}{p\textless0.001} \\
    \hline
    Cliff's Delta & 0.13  & \multicolumn{1}{c|}{-} & -0.50  & \multicolumn{1}{c|}{L} & 0.15  & \multicolumn{1}{c|}{S} & \multicolumn{1}{c|}{-} & \multicolumn{1}{c|}{-} & -0.53 & \multicolumn{1}{c|}{L} & -0.08 & \multicolumn{1}{c}{S} \\
    \hline
  \textbf{Smelly Dev.?} & \multicolumn{1}{c|}{\textbf{Y}} & \multicolumn{1}{c|}{\textbf{N}} &
\multicolumn{1}{c|}{\textbf{Y}} & \multicolumn{1}{c|}{\textbf{N}} &\multicolumn{1}{c|}{\textbf{Y}} & \multicolumn{1}{c|}{\textbf{N}} &\multicolumn{1}{c|}{\textbf{Y}} & \multicolumn{1}{c|}{\textbf{N}} &\multicolumn{1}{c|}{\textbf{Y}} & \multicolumn{1}{c|}{\textbf{N}} &\multicolumn{1}{c|}{\textbf{Y}} & \multicolumn{1}{c}{\textbf{N}} \\
    \hline
    Mean  & 0.27  & 0.33  & 0.03  & 0.19  & 0.21  & 0.20   & 0.58  & 0.57  & 0.03  & 0.13  & 1.00     & 1.03 \\
    \hline
    Variance & 0.08  & 0.24  & 0.00     & 0.13  & 0.01  & 0.02  & 0.02  & 0.06  & 0.01  & 0.06  & 0.03  & 0.10 \\
    \hline
    \end{tabular}%
  \label{tab:addlabel}%
\end{table*}%

\begin{figure*}[!htbp]
\centerline{\includegraphics[height=4.5cm ,width=14.5cm,angle=0]{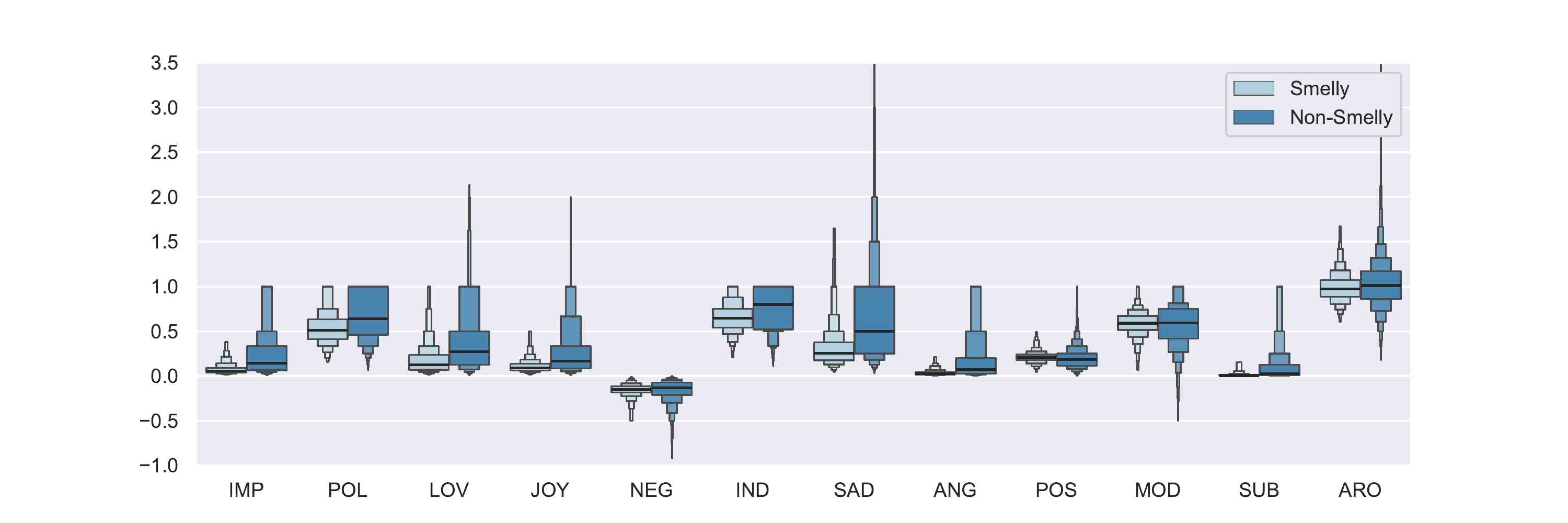}}
\caption{Distribution of the features.}
\label{fig}
\end{figure*}

\subsection{RQ3: Smelly vs. Non-Smelly Developers}

Fig. 8 depicts the distribution of sentimental features using an enhanced version of box-plot called boxen-plot \cite{boxenplot}, which is more capable of displaying tails of large-sampled data, as it cuts data into more quantiles. 

Table V lists the result of statistical analysis in order to differentiate the sentimental characteristics derived from our dataset, in which \{Y, N\} represent smelly and non-smelly Developers. Effect sizes in \{Large, Medium, Small, Negligible\} are mapped to \{L, M, S, -\}. Since we analyze the distribution of metrics' data separately, zeros are removed for each group of metrics in this section to ensure the effectiveness of data. 

Since predictive power could speak for the actual impact of features on the prediction outcome \cite{PalombaCSPredict, PalombaMSR}, we reveal IMP, IND, POL, SAD, LOV, and JOY may have impacts on community smell, which would be analyzed and discussed in the next paragraphs. 

The IMP feature is the top-ranked feature in terms of predictive power. From experience, we assume instructive expressions are more likely to be serious and impolite, which would become an obstacle for cooperation. However, result shows the distribution of the proportion of imperative GM is significantly different for smelly and non-smelly developers with a large effect size. Result also reveals non-smelly developers use 3 times more frequently the imperative expressions, \textit{i.e.}, commands and warnings.  Moreover, indicative GM is also among the top contributing features, and smelly developers use less indicative expressions than non-smelly ones. Hence, we conclude that ensuring certainty is vital to developers' communication and collaboration quality. This conclusion is in line with the observations made from purely technical aspects\cite{uncertainty}. Interestingly, non-smelly developers also use more subjunctive expressions to express wishes and opinions, indicating the importance of certainty may differ from scenarios. 

Difference in politeness, however, does not have a great effect size as we expect, \textit{i.e.}, in our dataset, smelly developers communicate in a similar structure compared with non-smelly developers in terms of politeness. Nevertheless, non-smelly developers use 24\% more frequently the polite expressions in general than smelly developers. 

Previous researches \cite{HAPPY_PULL_REQUEST_MERGED,graziotin2014happy} in developers' sentiments topics mainly regard happiness as a positive factor to development task. Surprisingly, expressing more joyful emotion may also indicate community smells' occurrence. Meanwhile, we also notice literature reporting that too many positive comments \cite{Huq} indicate potential bugs. In contrast, negative emotions have always been an indicator of potential problems in software engineering\cite{eeg}. Unexpectedly, except for anger, no notable difference are found in smelly and non-smelly developers in terms of negative sentiments. According to their mean values, non-smelly developers are even more bad-tempered and depressed than the smelly ones.  Due to the multi-facet and complex nature of sentiments and development activity, the difficulties in interpretation occurred frequently in developer sentiment analysis \cite{Ortu16MSRVAD,hawaii,linus,Cheruvelil19,Ortu15Politeness}. 

Practically, we suggest developers should communicate in a straightforward and polite way. As for emotions, we cannot provide suggestions until further empirical investigation and case studies are made to figure out positive and negative emotions' actual impact.  Indeed, the unforeseen part of results shed light on the necessity of a deeper understanding of developers' sentiments and their impact on software community through qualitative and quantitative study. Beyond technical aspects, the research community need to improve and reshape the framework of comprehending developers' perception\cite{PalombaMSR, CONTEXT_DEV_2,CONTEXT_DEV1,sensitive} as well as their task context\cite{psy, CONTEXT_BASED_REFACTORING_ICPC_2017,CONTEXT_2,CONTEXT_3}.

\begin{framed}
 Finding 3. Smelly developers are different from non-smelly ones in terms of sentiments. They are less polite, and they use less imperative and indicative GM, \textit{i.e.}, less certain statements, instructions and warnings. Unexpectedly, they express more positive and less negative emotions. To ensure community healthiness, we suggest developers should communicate in a straightforward and polite way.  
\end{framed}

\section{Threats to Validity}
This section clarifies the way we address threats to validity. 
\subsection{Construct Validity}
The major threat to construct validity is the reliability of our datasets. We combine 2 sources of information, \textit{i.e.}, community smell detection results and developer sentiment dataset.

In terms of community smell detection, we employ an open-source tool called \textsc{Codeface4Smells} \cite{PalombaCSDetect}. The authors provided detailed replication data to prove the dependability of the tool, \textit{i.e.}, its output were all true positives. Hence, we believe the tool is reliable. In addition, we follow strictly the installation, configuration, and execution guides\cite{PalombaMasters} of the detection tool. As for software repositories and mailing lists, we fetch them from original sources of \textsc{ASF} and \textsc{JBoss}. To a great extent, we can confirm the reliability of this source.

The developer sentiment dataset is proposed and improved progressively by Ortu \textit{et al.} \cite{Ortu16MSRDataset,Ortu14MSREmotions,Ortu15MSRBullies,Ortu15Politeness,Ortu16MSRVAD} through years of validated works. Except for the emotions (joy, love, anger, sadness), all data are automatically detected by lexicon-based tools. The performance of the tools may be a threat to validity. Indeed, no tool is ready for detecting sentiments in all kinds of discussions in software engineering\cite{cross-platform}, \textit{e.g.}, app reviews and Q\&As. However, tools employed by the dataset are all state-of-the-arts, \textit{e.g.}, \textsc{SentiStrength}\cite{SentiStrengthOrig} achieved an agreeable performance ranging from 0.70 to 0.99 in terms of positive and negative sentiment detection\cite{CHALLENGE_BUT_SENTISTRENGTH_HAS_RELIABLE_OUTPUT_ON_JIRA_ICSE18} in the \textsc{JIRA} ITS, and its reliability have also been proved in other researches\cite{gold,mlSentiStrength}. Thus, we conclude that the reliability of the sentiment dataset in the context of our research is acceptable.

The process of combining the two datasets, however, may cause some loss in information. To match the developers in both datasets, we make our best effort to link developers from both sides with their e-mails and names. However, 8.8\% of the smelly developers are not found in the sentiment dataset. Thus, their data are dropped. Nevertheless, we still manage to preserve most of the data.

The coherence of mailing lists and developer comments in \textsc{JIRA} may be a threat as well. Therefore, we investigate the contents of mailing lists. In most of the cases (8 out of 12), they are automatically generated \textsc{JIRA} discussions. If not so, they are \textsc{JIRA}-centered discussions, \textit{i.e.}, comments attaching hyperlinks of \textsc{JIRA} tickets. The involvement of ITS contents in mailing lists was also observed in other research \cite{mlcontent}. In conclusion, we are measuring discussions in the same context presented in different layouts.  

\subsection{Conclusion Validity}

The mix of manually labeled emotions and automatically detected sentiments in the dataset may hinder the practical value of our model, as the evaluation of sentiments is not fully automatic. However, such labels could be replaced by the outputs of reliable sentiment evaluation tools\cite{EmoTxt,emoji}.

In terms of the reliability of model settings, we configure the hyper-parameters using Grid Search, and we employ LOOCV as well as 10 $\times$ 10-Fold Validation, which was proved stable by previous research\cite{esd1}. We also report results of classical evaluation metrics, \textit{i.e.}, precision, recall, F-Measure, and AUC-ROC. Furthermore, we apply statistical tests, \textit{e.g.}, SK-ESD and Effect Sizes, to validate the significance of our conclusions. 

\subsection{External Validity}
Since different projects may involve developers in various backgrounds\cite{PalombaCSDiversity,novice}, the multi-faceted nature of software development and developer sentiment is an unavoidable threat to external validity. To address this issue, we perform our study in 12 active open-source projects of 2 major open source ecosystems to maximize the generality of our conclusion. Such systems have been widely studied in previous works of software engineering \cite{smellbugprediction,PalombaMSR,apachedebt}.

\section{Conclusion}

This paper investigates whether, and to what extent, the occurrence of community smells on developers could be predicted by their sentiments. Furthermore, it analyzes the difference of smelly and non-smelly developers in terms of the distribution and intensities of their sentiments. 

We construct sentimental features from a developer sentiment dataset\cite{Ortu16MSRDataset} as independent variables. Meanwhile, we exploit \textsc{Codeface4smells}\cite{PalombaMasters,PalombaCSDetect} to generate dependent variables concerning whether a developer is affected by 3 community smells, \textit{i.e.}, \textit{Lone Wolf}, \textit{Organization Silo}, and \textit{Bottleneck}, and if he/she is a \textit{Smelly Quitter}. Afterwards, we also assess the predictive power of all the features. Finally, we evaluate the significance and effect sizes of the distributions of sentiments on smelly and non-smelly developers. 

Result shows our model achieves mean F-Measures ranging from 76\% to 93\% in within- and cross-project prediction. Additionally, sentimental features including imperative and indicative GM, politeness, sadness, love, and joy are stronger predictors than the activeness metrics. We reveal smelly developers are less polite, and they use less certain statements, instructions, and warnings. Unexpectedly, we also discover that smelly developers express more positive and less  negative emotions, which need to be interpreted in further research. To conclude, we suggest developers should communicate in a straightforward and polite way. 

Future work includes: (1) the integration of more effort- and process-aware metrics, (2) involving other kinds of discussions such as chat messages\cite{chat}, (3) interpreting the pattern of positive and negative emotions' interaction with social debt.

\section*{Acknowledgment}

This work is partially supported by the NSF of China under grants No. 61772200, the  Shanghai Natural Science Foundation No. 17ZR1406900, 17ZR1429700, and the Software and Integrated Circuit Industry Development Special Funds of Shanghai Economic and Information Commission under Grant No. XX-XXFZ-02-20-2463.

\newpage

\bibliographystyle{ieeetr}
\bibliography{ref}

\end{document}